\documentstyle[eqsecnum,pre,preprint,aps,epsfig]{revtex}

\tightenlines

\newcommand{\bq}{\begin{equation}}
\newcommand{\eq}{\end{equation}}
\newcommand{\bqa}{\begin{eqnarray}}
\newcommand{\eqa}{\end{eqnarray}}
\newcommand{\ben}{\begin{enumerate}}
\newcommand{\een}{\end{enumerate}}
\newcommand{\bc}{\begin{center}}
\newcommand{\ec}{\end{center}}
\newcommand{\bqb}{\begin{eqnarray*}}
\newcommand{\eqb}{\end{eqnarray*}}

\begin{document}

\draft
\preprint{PM/04-01,~~corrected version}

\title{\vspace{1cm}  Supersymmetric virtual effects in heavy
quark pair production at LHC
\footnote{Partially supported by EU contract HPRN-CT-2000-00149}}
\author{M. Beccaria$^{a,b}$,
F.M. Renard$^c$ and C. Verzegnassi$^{d, e}$ \\
\vspace{0.4cm}
}

\address{
$^a$Dipartimento di Fisica, Universit\`a di
Lecce \\
Via Arnesano, 73100 Lecce, Italy.\\
\vspace{0.2cm}
$^b$INFN, Sezione di Lecce\\
\vspace{0.2cm}
$^c$ Physique
Math\'{e}matique et Th\'{e}orique, UMR 5825\\
Universit\'{e} Montpellier
II,  F-34095 Montpellier Cedex 5.\hspace{2.2cm}\\
\vspace{0.2cm}
$^d$
Dipartimento di Fisica Teorica, Universit\`a di Trieste, \\
Strada Costiera
 14, Miramare (Trieste) \\
\vspace{0.2cm}
$^e$ INFN, Sezione di Trieste\\
}

\maketitle

\begin{abstract}
We consider the production of heavy ($b,t$) quark pairs at proton colliders in 
the
theoretical framework of the MSSM. Under the assumption of a
"moderately" light SUSY scenario, we first compute the leading logarithmic
MSSM contributions at one loop for the elementary processes of
production from a quark and from a gluon pair in the 1 TeV c.m. energy region. 
We show that in the
initial gluon pair case (dominant in the chosen situation at LHC energies) the 
electroweak
and the strong SUSY contributions concur to produce an enhanced
effect whose relative value in the cross sections could reach the twenty
percent size for large $\tan\beta$ values in the realistic proton-proton LHC
process.

\end{abstract}
\pacs{PACS numbers: 12.15.-y, 12.15.Lk, 13.75.Cs, 14.80.Ly}

\section{Introduction}

One of the main goals of the future experiments at hadron colliders
will be undoubtedly the search for supersymmetric particles. In the
specific theoretical framework of the MSSM, a vast amount of
literature already exists showing the expected experimental
reaches for the different supersymmetric spectroscopies and
parameters, both for the TEVATRON\cite{tevatron} and for the
LHC\cite{LHC} cases. For what concerns the possibility of direct
production, nothing has to be added (in our opinion) to the existing
studies, leading to the conclusion that, if any supersymmetric
particle exists with a not unfairly large mass, it will not escape
direct detection and identification.\par
A nice and special feature of the present and future hadron colliders
is the fact that, on top of direct SUSY production, a complementary
precision test of the involved model (\underline{assuming} a
preliminary discovery) is also in principle possible. This implies the
study of virtual SUSY effects in the production of suitable (not
necessarily supersymmetric) final states, in full analogy with the
previous memorable analyses performed to test the Standard Model at 
LEP1, SLC and
LEP2 \cite{LEPSLC}. 
For what concerns the realistic experimental accuracy
requested by a similar search, one expects 
\cite{Moretti}, \cite{accLHC} a possible
relative few percent level. In this spirit, the existence of virtual
SUSY effects as large as a relative ten percent or more should be
carefully examined and investigated.\par
The aim of this preliminary paper is actually that of showing that the
production of a final heavy ($b,t$) quark pair at LHC could be particularly
convenient for the search of virtual SUSY effects. This is due to a
number of technical features (cancellations of disturbing
contributions at high energies, not negligible final quark masses,...)
that will be thoroughly illustrated in the following Sections. 
%As a
%matter of fact, we must say, though, that we would have expected
%a relevance of the role of the virtual corrections for final heavy
%quark pairs from our previous studies of electron-positron
%annihilation at LC\cite{top}. These were performed under the
%assumptions of a "relatively" light SUSY scenario, i.e. one where all
%the relevant masses of the process, for a c.m. energy assumed in the
%1 TeV range, were below, approximately, 350-400 GeV. Under such
%assumptions, a logarithmic expansion of Sudakov kind was used at one
%loop for the electroweak component of the process, showing a
%supersymmetric relative effect on the cross sections of a possible ten
%percent size (the calculation of the corresponding supersymmetric QCD
%effect was not carried on). Given these premises, it seemed natural to
%us to investigate the supersymmetric effects in an analogous situation
%at LHC where in principle a c.m. energy 
%in the 1 TeV range 
%can be reached. 
In this preliminary analysis, we studied the electroweak and the SUSY
QCD contributions to the subprocess initiated by $q\bar q$ and $gg$
partons and we limited our application to the invariant mass
distribution of $PP\to q\bar q+...$ at LHC
using standard quark and gluon structure functions. 
The rewarding result was that of finding that the SUSY
electroweak and the SUSY QCD terms concur to produce an overall effect
that can be as large as a relative twenty percent in the assumed light
SUSY scenario. This effect is more due to the electroweak component
than to the QCD component, essentially owing to the final heavy quark
Yukawa contribution (proportional to the quark squared mass). The
picture at one loop is in fact, surprisingly to us, 
practically the same that
one would find in the case of electron-positron annihilation at LC \cite{top},
as we shall briefly discuss  in this paper for sake of comparison.
At the possible few percent experimental accuracy level
\cite{accLHC}, this effect
should not escape a dedicated measurement, and could provide an
important test of the (assumedly -and hopefully- discovered)
supersymmetric scenario.\par
Technically speaking, this paper is organized as follows: Section 2
will be devoted to a description of the one-loop analysis for an
initial $q\bar q$ state; in Section 3, the same description will be
given for an initial $gg$ state and a brief comment will be given
on the analogy of the results with those obtained for an initial 
electron-positron state; in
Section 4, a numerical analysis of the possibly visible effects on the realistic
proton-proton initiated process will be shown and a few conclusions will finally 
appear in Section 5.

\section{Initial \lowercase{$q\bar q$} state}

We begin our analysis at the partonic level considering as initial
states the light quark components of the proton (including
the bottom component but ignoring the top quark one). 
At lowest $\alpha_s$ order the scattering amplitude is given by the
Born terms schematically represented in Fig.1. The $s$-channel diagram
of Fig.1a applies to all initial $q\bar q$ pair annihilation
 and leads to:

\bq
A^{\rm Born}_s=\sum_{i,j}a^{\rm Born~s}_{ij}[\bar v(\bar q)
{\lambda_l\gamma^{\mu}\over2}P_i u(q)][\bar u(q')
{\lambda_l\gamma_{\mu}\over2}P_j v(\bar q')]
\label{borns}\eq

\noindent
whereas the $t$-channel diagram will only apply to
$b\bar b\to b\bar b$ scattering 
(because we shall neglect the $t\bar t\to t\bar
t$ contribution) and writes:

\bq
A^{\rm Born}_t=\sum_{i,j}a^{\rm Born~t}_{ij}[\bar v(\bar q)
{\lambda_l\gamma^{\mu}\over2}P_i v(\bar q)][\bar u(q)
{\lambda_l\gamma_{\mu}\over2}P_j u(q)]
\label{bornt}\eq
\noindent
where $i,j$ refer to $R$ or $L$ quark chiralities
with $P_{R,L}={1\pm\gamma^5\over2}$;
$l$ is the intermediate gluon colour state and

\bq
a^{\rm Born~s}_{ij}={4\pi\alpha_s\over s}~~~~~~~~~~~~~~~
a^{\rm Born~t}_{ij}={4\pi\alpha_s\over t}
\eq

At the Born level, electroweak contributions only consist in
the Drell-Yan process $q\bar q \to\gamma , Z\to q'\bar q'$ but
the effect in the cross section will be reduced by a factor
$\alpha^2/\alpha^2_s\simeq0.01$ as compared to the QCD one
and will be neglected at the expected LHC accuracy.\par
Our starting assumption, as already stated in the Introduction, will
be the discovery of (at least some) supersymmetric particles. In this
spirit, we shall examine the possibility of performing a precision
test of the candidate model by looking at its higher order virtual
effects in the considered heavy quark production at hadron colliders,
starting from the determination of these effects at the partonic
level. As a first case to be examined, we shall restrict our study to
the MSSM framework. For what concerns the higher order diagrams to be
retained, we must at this point make our strategy completely clear. In
general, the higher order corrections to the considered processes will
be of electroweak and of strong nature. We shall begin our analysis
with the detailed study of the electroweak components given in the
following Section IIa.

\subsection{Electroweak corrections}

To the first order one-loop
level the electroweak corrections
 will typically correspond to the diagrams
schematically represented in Fig.2. Note that we considered in the
electroweak sector both the SM components and the genuinely
supersymmetric ones, that will be often denoted for simplicity with a
(SUSY) apex. In our analysis, we shall not consider higher order (e.g.
two loops) electroweak corrections. This (pragmatic) attitude is,
at least, 
supported by the
fact (that we showed in a previous paper\cite{resum}) that such terms
at lepton colliders (LC) are only requested if the 
available c.m. energy of the 
process is
beyond, roughly, the 1 TeV range. In our investigation of proton colliders, the 
structure of the leading
electroweak corrections is essentially similar to the LC one (with 
straightforward modifications of the
initial vertex contributions). To reproduce the benefit 
of the absence of hard 
resummation computations, we shall therefore
be limited to c.m. energy values not beyond the 1 TeV limit.\par
Having fixed the considered energy range, a numerical evaluation at
one loop can now be performed in a (reasonably) straightforward way.
A great simplification can nonetheless be achieved under the
assumption that the c.m. energy is sufficiently larger than all the
masses of the (real and virtual) particles involved in the process. In
this case, a logarithmic expansion of the so-called Sudakov kind can
be adopted. All the details of such an approach for the specific MSSM
model have been already exhaustively discussed in previous references
for an initial electron-positron state\cite{ee},\cite{validity}, 
but the results are
essentially identical if the initial state is a $q\bar q$ pair 
(as we said, one
must only modify in a straightforward way the initial vertex
contribution). To avoid loss of space and time, we defer thus the
reader to\cite{ee},\cite{validity} for details. 
The only relevant point that we
want to make is the fact that the "canonical" logarithmic expansion is
given to next-to leading order, i.e. retaining the double and the
linear logarithmic terms, and ignoring possible extra (e.g. constant)
contributions. To this level, the remarkable simplification for the
considered heavy quark final state is that only two SUSY parameters
appear in the expansion. One parameter is $\tan\beta$, contained in
the coefficient of the linear $ln$ of Yukawa origin; the second one is
a (common) SUSY mass scale $M$, to be understood as an "average"
supersymmetric mass involved in the process.\par
Given the fact that the supersymmetric masses might be not
particularly small (the existing limits for squark and gluino
masses are at the moment consistent with a lower bound of approximately $300$
GeV\cite{tevatron},\cite{limits}) and taking into account the qualitative 1 TeV
upper bound on the c.m. energy requested for the validity of a one-loop
expansion, we shall adopt in this preliminary analysis a reasonable
working compromise. In other words, we shall consider energy values in
the 1 TeV range, and a SUSY scenario "reasonably" light i.e. one in
which all the relevant SUSY masses of the process are smaller than,
approximately, $350-400$ GeV. This will encourage us to trust a
conventional Sudakov expansion, supported by previous detailed
numerical analyses given for electron-positron
accelerators\cite{ChargedHiggs}. 
%Before initiating the procedure, we would
%like, though, to make two general statements that seem to us
%appropriate. The first one concerns the fact that we shall only treat
%in this paper the SUSY contribution to next-to-leading logarithmic order, i.e.,
%we shall neglect e.g. possible constant \underline{SUSY} terms. These
%will depends in principle on many SUSY parameters and their estimate
%can be, in principle, performed once the values of the parameters are
%known, which for the moment seems to us beyond the purposes of this
%preliminary search at the logarithmic level.
%The second statement is that our analysis could be obviously performed
%in a less favorable situation i.e. one where the c.m. energy  were
%\underline{not} much larger than the relevant SUSY masses. The price
%to pay would be the abandon of the simple Sudakov expansion, which should not
%imply, in our opinion, the loss of validity of the one-loop expansion
%for the SUSY effect, as we shall try to discuss in the conclusions of
%this paper.\par

The previous discussion was related to the SUSY electroweak effects.
Our attitude will be in conclusion that of considering both the SM and
the SUSY components at the same one loop level. Strictly speaking,
since we are only interested in the SUSY effect, we could even ignore
the SM contribution. The reason why we retain it is simply that it is
easy to estimate it and interesting to show the "SUSY enhancement" in
the MSSM overall correction.\par 
After this long but, we hope, sufficiently clear discussion, we are
now ready to illustrate the one-loop numerical details. With this aim,
we shall start from the electroweak diagrams of Fig.2 and write,
following our previous notations \cite{resum,ee,validity}:\\

{\bf Electroweak logarithmic corrections on the amplitudes}\\

For each $q\bar q \to q'\bar q'$ amplitude we can write, following
ref.\cite{validity}:

\bq
A^{\rm 1~loop~ew}=A^{Born}[1+c^{\rm ew}]
\label{coef}\eq
\noindent
with
\bq
c^{\rm ew}\equiv c_{in, ~gauge}+c_{in, ~Yuk}+c_{fin~gauge}
+c_{fin, ~Yuk}+c_{ang}
\label{coefew}\eq
\noindent
where $c_{in}$ refers to the universal (i.e. process
independent) corrections due to the initial
$q\bar q$ pair, $c_{fin}$ to the universal corrections due to the 
final $b\bar b$ or $t\bar t$ pair and $c_{ang}$ is a process and
angular dependent correction typical of each $q\bar q \to q'\bar q'$
case. The Yukawa contributions are sizable only for $q$ (or
$q'$) being $b$ or
$t$ quarks. The coefficients $c^{\rm ew}$ receive 
SM and SUSY contributions, depending on the quark chirality.
The SM contributions are ($u$ represents both up and charmed quarks,
and $d$ both down and strange quarks):

\bqa
&&c^{u\bar u~SM}_{in,~gauge~L}=
c^{t\bar t~SM}_{fin,~gauge~L}={\alpha(1+26c^2_W)\over144\pi s^2_Wc^2_W}
[3ln{s\over M^2_W}-ln^2{s\over M^2_W}]
\nonumber\\
&&c^{u\bar u~SM}_{in,~gauge~R}=
c^{t\bar t~SM}_{fin,~gauge~R}={\alpha\over9\pi c^2_W}
[3ln{s\over M^2_W}-ln^2{s\over M^2_W}]
\eqa

\bqa
&&c^{d\bar d~SM}_{in,~gauge~L}=
c^{b\bar b~SM}_{fin,~gauge~L}={\alpha(1+26c^2_W)\over144\pi s^2_Wc^2_W}
[3ln{s\over M^2_W}-ln^2{s\over M^2_W}]
\nonumber\\
&&c^{d\bar d~SM}_{in,~gauge~R}=
c^{b\bar b~SM}_{fin,~gauge~R}={\alpha\over36\pi c^2_W}
[3ln{s\over M^2_W}-ln^2{s\over M^2_W}]
\eqa

\bqa
&&c^{b\bar b~SM}_{in,fin,~Yuk~L}=
c^{t\bar t~SM}_{in,fin,~Yuk~L}=-~{\alpha\over16\pi s^2_W}
[{m^2_t\over M^2_W}+{m^2_b\over M^2_W}][ln{s\over M^2_W}]
\nonumber\\
&&
c^{b\bar b~SM}_{in,fin,~Yuk~R}=-~{\alpha\over8\pi s^2_W}
[{m^2_b\over M^2_W}][ln{s\over M^2_W}]
\nonumber\\
&&c^{t\bar t~SM}_{in,fin,~Yuk~R}=-~{\alpha\over8\pi s^2_W}
[{m^2_t\over M^2_W}][ln{s\over M^2_W}]
\eqa

\bq
c^{q\bar q ~q'\bar q'~SM}_{ang ~ij}=-~{\alpha\over4\pi} 
[ln{1-cos\theta\over1+cos\theta}]
[8Q_qQ_{q'}+2{g^Z_{qi}g^Z_{fj}\over s^2_Wc^2_W}][ln{s\over M^2_W}]
\eq
This last angular dependent term refers to chirality states
$(ij)=(LL,LR,RL,RR)$ and  involves the quark charges $Q_q$ and
the $Zq\bar q$ couplings
$g^Z_{qL}=I^3_{qL}(2-4|Q_q|s^2_W)$, $g^Z_{qR}=-2Q_qs^2_W$.\\

The additional SUSY contributions affect only the universal parts:

\bqa
&&c^{u\bar u~SUSY}_{in,~gauge~L}=
c^{t\bar t~SUSY}_{fin,~gauge~L}=-~{\alpha(1+26c^2_W)\over144\pi s^2_Wc^2_W}
[ln{s\over M^2}]
\nonumber\\
&&c^{u\bar u~SUSY}_{in,~gauge~R}=
c^{t\bar t~SUSY}_{fin,~gauge~R}=-~{\alpha\over9\pi c^2_W}
[ln{s\over M^2}]
\eqa

\bqa
&&c^{d\bar d~SUSY}_{in,~gauge~L}=
c^{b\bar b~SUSY}_{fin,~gauge~L}=-~{\alpha(1+26c^2_W)\over144\pi s^2_Wc^2_W}
[ln{s\over M^2}]
\nonumber\\
&&c^{d\bar d~SUSY}_{in,~gauge~R}=
c^{b\bar b~SUSY}_{fin,~gauge~R}=-~{\alpha\over36\pi c^2_W}
[ln{s\over M^2}]
\eqa

\bqa
&&c^{b\bar b~SUSY}_{in,fin,~Yuk~L}=
c^{t\bar t~SUSY}_{in,fin,~Yuk~L}=-~{\alpha\over16\pi s^2_W}
[{m^2_t\over M^2_W}(1+2\cot^2\beta)
+{m^2_b\over M^2_W}(1+2\tan^2\beta)][ln{s\over M^2}]
\nonumber\\
&&
c^{b\bar b~SUSY}_{in,fin,~Yuk~R}=-~{\alpha\over8\pi s^2_W}
[{m^2_b\over M^2_W}(1+2\tan^2\beta)][ln{s\over M^2}]
\nonumber\\
&&c^{t\bar t~SUSY}_{in,fin,~Yuk~R}=-~{\alpha\over8\pi s^2_W}
[{m^2_t\over M^2_W}(1+2\cot^2\beta)][ln{s\over M^2}]
\label{finyuk}\eqa

One can see (as emphasized in ref.\cite{validity}) that
the total MSSM gauge correction is quite simply
obtained by replacing the
logarithmic SM factor $[3ln{s\over M^2_W}-ln^2{s\over M^2_W}]$
by $[2ln{s\over M^2_W}-ln^2{s\over M^2_W}]$
and the total MSSM Yukawa correction by replacing the SM
$m^2_t$ by $2m^2_t(1+\cot^2\beta)$ and 
$m^2_b$ by $2m^2_b(1+\tan^2\beta)$.\\

{\bf Results for angular distributions, averaged over initial, 
summed over final
polarizations}\\

We now list explicitely the complete MSSM results for each subprocess.
According to the rules given just above it is easy to separate
the pure SM and the additional SUSY parts. 
We first
consider the subprocesses involving only the annihilation
channel of Fig.1a and Fig.2, and write:

\bqa
{d\sigma^{1~loop}\over
dcos\theta}={d\sigma^{Born}\over
dcos\theta}+{\pi\alpha^2_s\over18s}\{~(1+\cos^2\theta)~[S]
+2\cos\theta~ [D]~\}
\eqa

with

\bqa
&&{d\sigma^{Born}(q\bar q \to q'\bar q')\over
dcos\theta}={\pi\alpha^2_s\over9s}(1+\cos^2\theta)
\eqa

and 

\newpage

\underline{For $u\bar u \to b\bar b$}

\bqa
S_{u\bar u b\bar b}&=&{\alpha\over72\pi s^2_Wc^2_W}
(54-32s^2_W)[2ln{s\over M^2_W}-ln^2{s\over M^2_W}]\nonumber\\
&&
-{\alpha \over4\pi s^2_W}[{m^2_t\over
M^2_W}(1+\cot^2\beta)+{3m^2_b\over M^2_W}(1+\tan^2\beta)]
[ln{s\over M^2_W}]
\nonumber\\
&&-~{\alpha(4s^2_W-9)\over18\pi s^2_Wc^2_W}
ln{1-\cos\theta\over1+\cos\theta}[ln{s\over M^2_W}]
\eqa

\bqa
D_{u\bar u b\bar b}&=&{\alpha \over2\pi s^2_Wc^2_W}
ln{1-\cos\theta\over1+\cos\theta}[ln{s\over M^2_W}]
\eqa

\underline{For $d\bar d\to b\bar b$}

\bqa
S_{d\bar d b\bar b}&=&{\alpha\over72\pi s^2_Wc^2_W}
(54-44s^2_W)[2ln{s\over M^2_W}-ln^2{s\over M^2_W}]\nonumber\\
&&
-~{\alpha \over4\pi s^2_W}[{m^2_t\over
M^2_W}(1+\cot^2\beta)+{3m^2_b\over M^2_W}(1+\tan^2\beta)]
[ln{s\over M^2_W}]
\nonumber\\
&&+{\alpha(8s^2_W-9) \over18\pi s^2_Wc^2_W}
ln{1-\cos\theta\over1+\cos\theta}[ln{s\over M^2_W}]
\eqa

\bqa
D_{d\bar d b\bar b}&=&-~{\alpha \over2\pi s^2_Wc^2_W}
ln{1-\cos\theta\over1+\cos\theta}[ln{s\over M^2_W}]
\eqa

\underline{For $u\bar u \to t\bar t$}

\bqa
S_{u\bar u t\bar t}&=&{\alpha\over36\pi s^2_Wc^2_W}
(27-10s^2_W)[2ln{s\over M^2_W}-ln^2{s\over M^2_W}]\nonumber\\
&&
-{\alpha \over4\pi s^2_W}[{3m^2_t\over
M^2_W}(1+\cot^2\beta)+{m^2_b\over M^2_W}(1+\tan^2\beta)]
[ln{s\over M^2_W}]
\nonumber\\
&&-~{\alpha(16s^2_W+9) \over18\pi s^2_Wc^2_W}
ln{1-\cos\theta\over1+\cos\theta}[ln{s\over M^2_W}]
\eqa

\bqa
D_{u\bar u t\bar t}&=&-~{\alpha \over2\pi s^2_Wc^2_W}
ln{1-\cos\theta\over1+\cos\theta}[ln{s\over M^2_W}]
\eqa

\underline{For $d\bar d\to t\bar t$}

\bqa
S_{d\bar d t\bar t}&=&{\alpha\over36\pi s^2_Wc^2_W}
(27-16s^2_W)[2ln{s\over M^2_W}-ln^2{s\over M^2_W}]\nonumber\\
&&
-~{\alpha \over4\pi s^2_W}[{3m^2_t\over
M^2_W}(1+\cot^2\beta)+{m^2_b\over M^2_W}(1+\tan^2\beta)]
[ln{s\over M^2_W}]
\nonumber\\
&&-{\alpha(4s^2_W-9) \over18\pi s^2_Wc^2_W}
ln{1-\cos\theta\over1+\cos\theta}[ln{s\over M^2_W}]
\eqa

\bqa
D_{d\bar d t\bar t}&=&~{\alpha \over2\pi s^2_Wc^2_W}
ln{1-\cos\theta\over1+\cos\theta}[ln{s\over M^2_W}]
\eqa

\underline{For $b\bar b\to t\bar t$}

\bqa
S_{b\bar b t\bar t}&=&{\alpha\over36\pi s^2_Wc^2_W}
(27-16s^2_W)[2ln{s\over M^2_W}-ln^2{s\over M^2_W}]\nonumber\\
&&
-~{\alpha \over4\pi s^2_W}[{4m^2_t\over
M^2_W}(1+\cot^2\beta)+{4m^2_b\over M^2_W}(1+\tan^2\beta)]
[ln{s\over M^2_W}]
\nonumber\\
&&-{\alpha(4s^2_W-9) \over18\pi s^2_Wc^2_W}
ln{1-\cos\theta\over1+\cos\theta}[ln{s\over M^2_W}]
\eqa

\bqa
D_{b\bar b t\bar t}&=&~{\alpha \over2\pi s^2_Wc^2_W}
ln{1-\cos\theta\over1+\cos\theta}[ln{s\over M^2_W}]
\eqa

\underline{In the special case of $b\bar b \to b\bar b$}
we have to add the annihilation amplitude of Fig.1a and of Fig.2,
and the scattering amplitude of Fig.1b and of
the $s\to t$ crossed diagrams of Fig.2.
This leads to the result

\bqa
{d\sigma^{1~loop}(b\bar b \to b\bar b)\over dcos\theta}&=&
{d\sigma^{Born}(b\bar b \to b\bar b)\over dcos\theta}
\{~1+ {\alpha\over72\pi s^2_Wc^2_W}(27-22s^2_W)
[2ln{s\over M^2_W}-ln^2{s\over M^2_W}]\nonumber\\
&&-~{\alpha \over4\pi s^2_W}[{m^2_t\over M^2_W}
(1+\cot^2\beta)+{3m^2_b\over M^2_W}(1+\tan^2\beta)]
[ln{s\over M^2_W}]~\}\nonumber\\
&&-~{\alpha^2_s\alpha\over18s}
[ln{s\over M^2_W}]~\{~({18-8s^2_W\over9s^2_Wc^2_W})
(~[{u^2\over s^2}-~{u^2\over3st}]ln{1-\cos\theta\over1+\cos\theta}
\nonumber\\
&&
-[{u^2\over t^2}-~{u^2\over3st}]ln{1+\cos\theta\over2})~
+{27-22s^2_W\over9s^2_Wc^2_W}[{u^2+s^2\over t^2}-~{u^2\over3st}]
ln{1-\cos\theta\over2}
\nonumber\\
&&-~({4s^2_W\over9s^2_Wc^2_W})(~{2t^2\over s^2}
ln{1-\cos\theta\over1+\cos\theta}
-~{2s^2\over t^2}~
ln{1+\cos\theta\over2}~)~\}
\label{bbbbew}\eqa
with
\bqa
&&{d\sigma^{Born}(b\bar b \to b\bar b)\over
dcos\theta}={2\pi\alpha^2_s\over9s}~ [ {u^2+t^2\over s^2}
+{u^2+s^2\over t^2}-{2u^2\over3st}]
\eqa

In the expression  (\ref{bbbbew}), the first part
(which factorizes the Born term) is the universal effect
including the gauge term and the double Yukawa effect,
whereas the second part is the
angular dependent
effect from s and from t channels (one can check, restricting
to the $1/s^2$ terms, that one recovers the previous $d\bar d\to b\bar
b$ case); the $s\to t$
crossing relation between annihilation and scattering contributions
is also clearly satisfied.\\

Having completed the discussion of the electroweak SUSY effect, we now
move in the following subsection IIb to the discussion of the strong
(QCD) SUSY correction.

\subsection{QCD SUSY correction}

A preliminary statement to be made before entering the discussion of
the QCD SUSY corrections is that we shall treat this effect under the
same assumptions that we adopted for the electroweak sector,
subsection IIa. In other words, we shall still concentrate our
analysis on c.m. energy values in the 1 TeV region, assuming the
previously considered "reasonably" light SUSY scenario. This will
allow us to use the same kind of simple logarithmic expansion that was
exploited in the electroweak case, at the one loop level.\par
For what concerns the expected validity of a one-loop perturbative
expansion, the situation is now, though, drastically different from
that of subsection IIa, and requires a precise subtle choice of
strategy. It is actually well-known\cite{QCDcoll} that, in the SM 
domain, the one-loop QCD correction is not accurate enough and higher
order terms seem to be fundamental. This is understandable in an
extremely simplified fashion as a consequence of the small scale which
enters the SM running of $\alpha_s$. In this qualitative picture, one
would expect that the corresponding SUSY effects do not share this
dramatic problem, owing to the much larger mass scale involved. This
would justify the  expectation that, for the restricted subset of SUSY
QCD corrections, a one-loop calculation can be sufficient.\par
Another essential difference between the SM QCD and the SUSY
QCD corrections is that in the SM part the infrared logarithms
arising from virtual gluon
contributions cancel against those occurring in 
soft real gluon emission, leaving only "constant" terms. In the SUSY
case, large logarithms of virtual origin and scaled
by the average SUSY mass will remain.  
A practical
attitude seems to be therefore that of isolating the SM higher order
QCD effects, considering them as a "known" quantity, much in analogy
with what is done for the canonical QED corrections, and to factorize
an explicit one-loop term containing the genuine SUSY QCD correction,
to be added in the usual one-loop philosophy to the electroweak one.
This is the approach that we shall follow in the rest of the paper,
which is, as we said, only interested in the evaluation of the
\underline{genuine} overall SUSY effect.\par
Having made this preliminary statement, we move now to the evaluation
of the QCD SUSY effects at one loop. These are represented
schematically in Fig.3, and can be classified in two quite different
sectors, respectively of vertex kind and of RG origin. The first ones
are essentially similar to analogous electroweak vertices of Sudakov
kind, with a simple replacement of gauginos by gluinos. They produce,
in the adopted "Sudakov regime", the following effect on the
amplitude for each external $q\bar q$ pair

\bq
c^{q\bar q~SUSY~QCD}_{L,R}=-~{\alpha_s\over3\pi}ln{s\over M^2}
\label{coefqcd}\eq
\noindent
which leads to an additional coefficient in the series
eq.(\ref{coef},\ref{coefew})
equal to twice this value for a $q\bar q\to q'\bar q'$
annihilation amplitude or for a $q\bar q\to q\bar q$ 
scattering amplitude, i.e. eq.(\ref{coef}) is replaced by

\bq
A^{Born}[1+c^{\rm ew}
-~{2\alpha_s\over3\pi}ln{s\over M^2}]
\label{coefewqcd}\eq
\noindent
From eq.(\ref{coefewqcd}) a rather important (in our opinion) 
feature can be
stressed. Numerically, one sees that for "reasonable" values
of the squark and gluino masses, the QCD SUSY vertex effect has a
numerical value of approximately 5 percent at 1 TeV c.m. energy. This
supports our assumption that higher order terms can be neglected, as
we shall do in this paper. Note also that the size of the effect is
smaller than that of the analogous electroweak component, particularly
for large $\tan\beta$ values in the Yukawa couplings. In
other words, at the one loop level, electroweak SUSY corrections
appear to be "stronger" than the corresponding QCD ones, in full
analogy with a similar feature first stressed in the electron-positron
case\cite{Ciafaloni}.\par
From a practical point of view, the very welcome feature that
characterizes the QCD SUSY vertex is the fact that the sign of the
one-loop effect is \underline{the same} (negative) as that of the
corresponding electroweak one. As a consequence of this rewarding
concurrence, the overall SUSY vertex effect at one loop gets an
enhancement that will improve the possibility of experimental
detection, and we shall return to this point in the final
discussion.\par
The next term to be computed corresponds to the RG diagram of Fig.(3e).
Following our previous discussion, we shall only consider the genuine
SUSY effect on the intermediate gluon bubble. This is given at one
loop from  standard formulae\cite{QCDcoll}. In a super-simplified
attitude of considering the relevant c.m. energy beyond \underline{all}
SUSY scales, it would lead to the following modification:

\bq
\delta \alpha_s^{SUSY}(s) = \alpha_s(M^2)
[-B_{SUSY}{\alpha_s(M^2)\over2\pi}ln{s\over M^2}]
\eq
The SUSY contribution arising in the MSSM for $s>M^2$ is
obtained using $B_{SUSY}=-2$.
This leads effectively to one more additional coefficient 
in the series of
corrections eq.(\ref{coef},\ref{coefew}) to the annihilation amplitude

\bq
c^{q\bar q, SUSY QCD}_{RG}=
-B_{SUSY}{\alpha_s(M^2)\over2\pi}ln{s\over M^2}
\label{coefrg}\eq
\noindent
and a similar correction to the scattering amplitude with
$ln(s/M^2)$ replaced by $ln(-t/M^2)$.\par

Note that, in a less optimistic situation of larger SUSY masses, the
effect would be reduced and should be computed more carefully. But,
independently of this, we notice a less pleasant feature 
of eq.(\ref{coefrg})
compared to eq.(\ref{coefewqcd}) i.e. the fact 
that it produces an effect of opposite
sign with respect to that of the electroweak one, 
thus reducing the overall SUSY
correction:

\bq
A^{\rm 1~loop~SUSY}=A^{Born}[c^{\rm SUSY~ew}+c^{\rm SUSY~QCD}]
\label{coefewqcdt}\eq

with 
\bq
c^{\rm SUSY~QCD}\simeq-~{2\alpha_s\over3\pi}ln{s\over M^2}
+{\alpha_s\over\pi}ln{s\over
M^2}={\alpha_s\over3\pi}ln{s\over M^2}
\eq

 From this point of view, the $q\bar q$ initial state
exhibits a "disturbing" feature for the detection of SUSY effects.
This feature will \underline{not} affect the determination of the SUSY
correction for the initial gluon gluon state at high c.m. energies, as
we shall show in the following Section 3 which will be devoted to the
study of that process.

\section{Initial \lowercase{gg} state}

The amplitude for the process $g^ig^j\to q'\bar q'$ 
(where $i,j$ denote the gluon color states)
is obtained by summing the 2 diagrams
of Fig.4a,b and the crossed diagram of Fig.4b:

\bq
A^{Born}_s= -i~{g^2_s\over s}f^{ijl}(\epsilon^i.\epsilon^j)
[\bar u(q'){\lambda^l\over2}\gamma^{\mu}(k^i-k^j)_{\mu} v(\bar q')]
\eq

\bq
A^{Born}_t= -~{g^2_s\over t}
[\bar
u(q'){\lambda^i\lambda^j\over4}(\gamma^{\mu}\epsilon^i_{\mu})
(\gamma^{\rho}(k^i-p^{q'})_{\rho})(\gamma^{\nu}\epsilon^j_{\nu})
 v(\bar q')]
\eq

\bq
A^{Born}_u= -~{g^2_s\over u}
[\bar
u(q'){\lambda^j\lambda^i\over4}(\gamma^{\mu}\epsilon^j_{\mu})
(\gamma^{\rho}(k^i-p^{\bar q'})_{\rho})(\gamma^{\nu}\epsilon^i_{\nu})
 v(\bar q')]
\eq
where $s=(k^i+k^j)^2=(p^{q'}+p^{\bar q'})^2$, $t=(k^i-p^{q'})^2$, 
$u=(k^i-p^{\bar q'})^2$, and $(\epsilon^i,k^i)$, $(\epsilon^j,k^j)$
are the polarization vectors and four-momenta of the gluons.

It is important to notice the gauge cancellations occurring at
high energies. From the above expressions one can immediately
compute the helicity amplitudes 
$F(\tau^i,\tau^j,\lambda^{q'},\lambda^{\bar q'})$ with 
$\tau^i=\pm1$, $\tau^j=\pm1$, $\lambda^{q'}=\pm1/2$, 
$\lambda^{\bar q'}=\pm1/2$ 
being the gluons and quark helicities.
At high energy, neglecting quark masses, one obtains only
chirality conserving terms with 
$\lambda^{q'}=-\lambda^{\bar q'}\equiv\lambda=\pm1/2$.
The contribution to the amplitudes
with $\tau^i=-\tau^j\equiv\tau=\pm1$ arises from $t$ and
$u$ channel terms:

\bq
F^{Born}(\tau,-\tau,\lambda,-\lambda)
=g^2_s({\lambda^i\lambda^j\over4})
{2\lambda\cos\theta+\tau\over1-\cos\theta}\sin\theta
+g^2_s({\lambda^j\lambda^i\over4})
{2\lambda\cos\theta+\tau\over1+\cos\theta}\sin\theta
\eq
\noindent
whereas the $\tau^i=\tau^j\equiv\tau=\pm1$ amplitudes 
get contributions from $s$, $t$ and
$u$ channel terms:

\bq
F^{Born}(\tau,\tau,\lambda,-\lambda)
=
-ig^2_sf^{ijk}{\lambda^k\over2}(2\lambda)\sin\theta
+g^2_s({\lambda^i\lambda^j\over4})(2\lambda)\sin\theta
-g^2_s({\lambda^j\lambda^i\over4})(2\lambda)\sin\theta
\eq
and totally cancel as
${\lambda^i\lambda^j\over4}-{\lambda^j\lambda^i\over4}
=if^{ijk}{\lambda^k\over2}$.

So, at high energy,  we are only left with contributions to 
$F^{Born}(\tau,-\tau,\lambda,-\lambda)$ 
arising from the $t$ and $u$ channel
quark exchange diagrams, $\lambda=\pm1/2$ corresponding to $R,L$
chiralities respectively.

The electroweak corrections are then extremely simple.
At first order
no electroweak correction arises for gluons. Only the
universal gauge and Yukawa terms appear for
the final $b$ or $t$ quark pair: 
$c^{q'\bar q'}_{fin, ~gauge~L,R}$ and $c^{q'\bar q'}_{fin, Yuk~L,R}$
from eq.(\ref{coefew}).

The SUSY QCD corrections turn out to be also extremely simple.
Because of the gauge cancellation of the $s$ channel term,
at first order we can ignore the SUSY QCD corrections to this part
and only consider the $t$ and $u$ channel
quark exchange diagrams, Fig.1b and Fig.5. There is no SUSY QCD
correction to the external gluon lines because of the cancellation
between the gluon splitting function and the gluon coupling Parameter
Renormalization (this fact is similar to the one occurring for
electroweak gauge bosons as noticed in ref.\cite{Denner,WW}).
Only the universal SUSY QCD correction to the external quark lines
appear, given by $c^{q'\bar q'~SUSY~QCD}_{L,R}$
of eq.(\ref{coefqcd}).

The angular distributions averaged over initial $gg$, 
summed over final $b, \bar b$ or $t, \bar t$
polarizations are then given by

\bqa
&&{d\sigma^{1~loop}(gg\to b\bar b)\over
dcos\theta}={d\sigma^{Born}(gg\to b\bar b)\over
dcos\theta}~\{~1+{\alpha\over144\pi s^2_Wc^2_W}
(27-22s^2_W)[2ln{s\over M^2_W}-ln^2{s\over M^2_W}]\nonumber\\
&&-~{\alpha \over8\pi s^2_W}[{m^2_t\over M^2_W}
(1+cot^2\beta)+{3m^2_b\over M^2_W}(1+tan^2\beta)][ln{s\over M^2_W}]
-~{2\alpha_s\over3\pi}[ln{s\over M^2}]~\}
\label{ggbb}\eqa

\bqa
&&{d\sigma^{1~loop}(gg\to t\bar t)\over
dcos\theta}={d\sigma^{Born}(gg\to t\bar t)\over
dcos\theta}~\{~1+{\alpha\over144\pi s^2_Wc^2_W}
(27-10s^2_W)[2ln{s\over M^2_W}-ln^2{s\over M^2_W}]\nonumber\\
&&-~{\alpha \over8\pi s^2_W}[{3m^2_t\over M^2_W}
(1+cot^2\beta)+{m^2_b\over M^2_W}(1+tan^2\beta)][ln{s\over M^2_W}]
-~{2\alpha_s\over3\pi}[ln{s\over M^2}]~\}
\label{ggtt}\eqa

In the above expression we have written the complete MSSM 
electroweak correction and the SUSY QCD correction. In the MSSM   
electroweak part one can easily separate the pure
SM and the SUSY components using the rules already stated in the previous
section. In particular, the SUSY modification of the universal terms
consists of replacing the linear SM $\simeq 3ln$ term by $2ln$, thus
leading to a negative $-ln$ effect, while the Yukawa term produces a
substantial negative contribution due to the $\tan\beta$ parameter.\par

The Born part for $q'=b$ or $t$ is given by

\bq
{d\sigma^{Born}(gg\to q'\bar q')\over
dcos\theta}={\pi\alpha^2_s\over4s}[ {u^2+t^2\over 3ut}-
{3(u^2+t^2)\over 4s^2}]
\eq
\noindent
and in a complete computation the SM QCD corrections will
have to be added.\par
As one sees from eq.(\ref{ggbb},\ref{ggtt}), 
as a "technical" consequence of the $t,u$
channel diagrams dominance in the chosen configuration, the SUSY
electroweak and QCD effects concur at their (expectedly accurate)
one-loop level to produce an overall negative effect that can be
sizable, particularly for large $\tan\beta$ values where it could
reach a relative twenty percent size, as we shall show in detail in
the following Section. In this sense, and within the special
scenario of large c.m. energies and "reasonably" light SUSY masses
that we have fixed in this simplified preliminary analysis, the
chances of detecting a virtual SUSY effect in heavy quark production
appear thus to be more promising for an experimental situation such
that the considered initial gluon-gluon state gives (via its t and u
channel diagrams) the dominant contributions to the cross section.
From the available \cite{lumi} luminosity
pictures, we deduce that this request
indicates the LHC experiments. Therefore, from now on we shall
concentrate our investigation on this special machine, keeping in mind
that for a different scenario, whose analysis might be performed in a
less simple way, the role of TEVATRON might be relevant, or dominant,
as well.\par
As a final comment to this Section, we would like to remark the 
fact that the overall \underline{relative} "genuine"
SUSY correction at one loop, in the chosen LHC scenario, is 
almost identical with that 
which one one would find,
for the same heavy quark pair production in the identical c.m.
energy and SUSY scenario, at a lepton
collider (LC). Without writing additional explicit formulae, 
we can simply explain this statement with the
observation that, in the relevant gluon-gluon initiated process, 
the surviving SUSY effect is of purely
universal (i.e. process independent) kind and due to the
heavy quark final state, both for the electroweak 
and for the QCD SUSY components, the
latter ones being entirely of vertex kind owing to the previously 
shown RG suppression. These
contributions factorize in the same way and therefore produce the 
same relative SUSY effect for initial
gluon-gluon and electron-positron state, even if the relative 
Standard Model corrections may be different
for the two cases e.g. when non universal angular dependent 
terms appear. The only (small) differences are due to the
universal (non Yukawa) SUSY contribution (i.e. the $-ln(s/M^2)$
term) arising from the initial $e^+e^-$ state.\par 
Our analysis of the simple partonic processes is thus completed. 
The next step is now that of considering
to which amount the features that we have underlined will 
survive in the real process of production from a
proton-proton state. This will be done in the forthcoming Section.\par

\section{Cross section for heavy quark pair production 
in $PP$ collisions}

We now consider proton-proton collisions with inclusive production of 
a pair of heavy quarks $PP\to q'\bar q' + .....$ . In this preliminary
analysis we just want to show the role of the SUSY
corrections on both quark-antiquark and gluon-gluon subprocesses.
With this purpose we will concentrate our attention on
the invariant
mass distributions of final $b\bar b$ or $t\bar t$ quarks. Future 
works may consider other
types of distributions (like $p_T$ distributions of the quarks or
of their decay products) using
the subprocess amplitudes that we have established in Sect.II and
III, and the corresponding parton model kinematical tools.\par
 For a total c.m. squared energy
$S$, the $q'\bar q'$ squared invariant mass ($s$)  distribution 
is given by

\bqa
{d\sigma(PP\to q'\bar q'+...)\over ds}&=&
{1\over S}~\int^{\cos\theta_{max}}_{\cos\theta_{min}}
d\cos\theta~[~\sum_{ij}~L_{ij}(\tau, \cos\theta)
{d\sigma_{ij\to  q'\bar q'}\over d\cos\theta}(s)~]
\eqa
\noindent
where $\tau={s\over S}$, and $(ij)$ represent 
all initial $q\bar q$ pairs with 
$q=u,d,s,c,b$ and the initial $gg$ pairs, with the corresponding
luminosities

\bq
L_{ij}(\tau, \cos\theta)={1\over1+\delta_{ij}}
\int^{\bar y_{max}}_{\bar y_{min}}d\bar y~ 
~[~ i(x) j({\tau\over x})+j(x)i({\tau\over x})~]
\eq
\noindent
$i(x)$ being the distributions of the parton $i$ inside the proton
with a momentum fraction,
$x={\sqrt{s\over S}}~e^{\bar y}$, related to the rapidity
$\bar y$ of the $q'\bar q'$ system.
The limits of integrations for $\bar y$ can be written

\bqa
&&\bar y_{max}=\max\{0, \min\{Y-{1\over2}ln\chi,~Y+{1\over2}ln\chi,
~-ln(\sqrt{\tau})\}\}\nonumber\\
&&
\bar y_{min}= - \bar y_{max}
\eqa
\noindent
where the maximal rapidity is $Y=2$, the  
quantity $\chi$ is related to the scattering angle
in the $q'\bar q'$ c.m.
\bq
\chi={1+\cos\theta\over1-\cos\theta} 
\eq
and 
\bq
\cos\theta_{min,max}=\mp\sqrt{1-{4p^2_{T,min}\over s}}
\eq
expressed in terms of
the chosen value for $p_{T,min}$.\\

We have evaluated numerically the above expression of the differential cross section
in the LHC case with $\sqrt{S} = 14$ TeV, a fixed SUSY mass scale $M_{SUSY} = 350$ GeV and 
an angular cut corresponding to $p_{T, min} = 10$ GeV. 
Concerning the parton distributions, we must stress that in principle
we should consider their evolution up to the desired energy scale $\sqrt{s}$ in the 
framework of SUSY QCD. However, the supersymmetric corrections to the evolution 
should be negligible in a first approximate treatment if the masses of the supersymmetric 
particles are large enough, still not spoiling the applicability of the Sudakov expansion.
With these remarks in mind, we have used the 2003 NNLO MRST set of evolved parton distribution functions 
available on~\cite{lumi}.

A summary of our numerical analysis is shown in 
Figs.~(\ref{run1}-\ref{run3}).
In Fig.~(\ref{run1}), we show the percentual 
effect on the cross sections for production of 
final bottom or top quarks at two representative 
values $\tan\beta = 10, 40$
in the c.m. energy range 0.7-1 TeV (the lower limit 
corresponds, qualitatively, to a value
$s=4M_{SUSY}^2$ where we can still hope from our experience
in $H^+H^-$ study \cite{ChargedHiggs} that 
our logarithmic expansion is ``reasonable''). 
For the higher value, 
the effect reaches the remarkable 20 \% level. 
An analysis of the relative weights of the various subprocesses
contributing the total cross section shows that 
the dominant subprocess is the $gg$ one.
This is due to the low values of the ratio $s/S$ 
at LHC in the considered range for the final state invariant mass.
Indeed, at low $s/S$ the fraction $x$ is also 
typically small and the rapid rise of the gluon distribution 
function overwhelms the role of the other subprocesses; see for
example the illustrations for $gg$ and $q\bar q$ luminosities
given in ref.\cite{QCDcoll}.
To give some numerical examples one can check that 
for $\sqrt{s}$ = 1 TeV 
the contribution of the $gg$ subprocess represents 
about $80\%\ (88 \%)$ of the total Born cross section 
for final bottom (top). At the smaller $\sqrt{s}$ = 700 GeV, 
the effect is even larger with the $gg$
subprocess being now $85\%\ (92 \%)$ for final bottom (top).
Since the process is dominated by the $gg$ subprocess, 
the features of Figs.~(\ref{run1}-\ref{run3})
can be explained in terms of Eqs.~(\ref{ggbb},\ref{ggtt}).
In particular at the special value $\tan\beta=40$ the two 
Yukawa combinations proportional to $m_t^2$ and $m_b^2$
appearing in Eqs.~(\ref{ggbb},\ref{ggtt}) happen to be 
equal explaining the almost superposed lines in the plot~\footnote{
We used in this preliminary analysis the fixed values 
$m_t = 173.8$ GeV, $m_b$ = 4.25 GeV, rather than (energy dependent) {\em running} values. In so doing, 
we ignored a higher order effect consistently with the philosophy of our paper.}.

In Fig.~(\ref{run2}), we show the comparison between the 
effects in the Standard Model
and those that we find in the MSSM at $\tan\beta = 40$, 
a value that has the "advantage" of 
providing a large correction due to the Yukawa terms. 
Indeed, it is precisely this kind of contribution
that is responsible for
the significant enhancement of the effect compared with the Standard Model case. The reason is the 
amplification of the coefficient of the $m_t^2$ term by more than a factor 2 due to the replacement 
$m_t^2\to 2m_t^2(1+\cot^2\beta)$ as well as the additional large correction $\tan^2\beta$ introduced 
by the analogous replacement $m_b^2\to 2m_b^2(1+\tan^2\beta)$. As a comment about the numerics we 
remark that in the Standard Model the Yukawa effect for final top is larger than in the case of 
final bottom by the factor $(3m_t^2 + m_b^2)/(m_t^2 + m_b^2)\simeq 3$ explaining why the
Standard Model full effect is larger for final top. In the MSSM at $\tan\beta\simeq 40$ we already discussed
the equivalence of the Yukawa effect for the final top or bottom.

Finally, in Fig.~(\ref{run3}), we show the relative 
weights of the genuine SUSY contributions
that are not present in the Standard Model. 
They are three, {\em i.e.} (i) the SUSY component of the QCD 
correction (we already mentioned the genuine 
Standard Model QCD correction and we simply remark that is shared by 
both the Standard Model and the MSSM), (ii) 
the SUSY gauge electroweak Sudakov (linear) logarithmic term, 
(iii) the additional $\tan\beta$ dependent Yukawa terms. 
The dominant effects are the Yukawa and the SUSY QCD correction,
 the first one being, as anticipated in the 
Introduction, the largest. In agreement with the 
previous Figure (\ref{run2}), 
the Yukawa part of the difference ``MSSM - SM'' 
is larger for final bottom (at $\tan\beta = 40$). 
Indeed, as we remarked, we have MSSM(top) $\simeq$ MSSM(bottom) and 
SM(top) $>$ SM(bottom) leading to MSSM-SM(top) $<$ MSSM-SM(bottom). 
Notice also that minor numerical differences 
between the curves for final bottom or top 
must be traced back to the fact 
that for final bottom the cross section contains a small, 
but non negligible, component from the subprocess
$b\bar{b}\to b\bar{b}$ whose Born angular dependence 
is totally different than the counterpart 
$b\bar{b}\to t\bar{t}$ in the case of final top.

Figures~(\ref{run1}-\ref{run3}) show the main result of our paper. 
One sees that the relative overall SUSY effect could be large, 
varying
from approximately ten percent to approximately twenty percent 
in the range $\tan\beta=10-40$. This effect is
definitely larger than the corresponding SM one, 
as shown by Fig.~(\ref{run2}). In particular, one 
notices that the enhancement
is less due to the SUSY QCD contribution, and is 
mostly coming from the SUSY Yukawa term, which would 
be absent in the case of
light quark production. For an experimental and theoretical precision 
at the few percent level (i.e. summing statistics, detection
efficiencies and uncertainties in quark and gluon distributions), 
the presence of such a SUSY
correction represents therefore, in our opinion, 
a feature of the process that cannot be ignored, and could 
provide a rather stringent test
of the Supersymmetric model to be investigated. 
In this sense, the production of heavy quark pairs at LHC appears to
us to be particularly interesting.\par

\newpage

\section{Conclusions}

A number of realistic statements must be made when drawing 
some possible conclusions from this paper. Our analysis has 
undoubtedly been specific, since it has assumed a combination 
of events i.e. a previous discovery of Supersymmetric particles
and a "reasonably light " nature of the SUSY scenario. 
In this particular case, we have shown that in the production of
heavy quark pairs at c.m. energies in the one TeV range, 
where quite reasonably a one-loop logarithmic Sudakov expansion
should provide an accurate description of the virtual SUSY 
MSSM correction,
the latter could reach values in the
twenty percent range for large $\tan\beta$ and might be 
therefore detectable at proton colliders, in particular at the LHC
which seems to us to be, for the specific c.m. energy 
configuration that we have chosen, the more suitable machine.  Of the
relevant logarithmic expansion we have computed  the leading
(quadratic) and next
to leading (linear)term, the SUSY genuine effect being simply of linear kind. We have performed this computation leaving
undetermined a possible next-to-next-to leading term, in practice a constant one. In a complete treatment, the latter
term should be computed or at least estimated. This is not trivial since in this quantity a large number of parameters of the
supersymmetric model will generally appear, and several extra assumptions should be made concerning their values that
are, we believe, beyond the realistic purposes of this first preliminary 
analysis~\footnote{In particular, given the fact that all the logarithmic SUSY effects at one loop are of 
linear order, possible SUSY masses $M_i$ larger than $M_{SUSY}$ (e.g. masses of heavy gluinos) are 
automatically reabsorbed by constant terms $\sim \alpha\log M_i/M_{SUSY}$ that should though, remain tolerably
small, given the absence of ``Yukawa enhancement''.}. 
%We expect from our previous experience
%in the case of electron-positron colliders~\cite{ChargedHiggs,BProduction} that such extra constant terms do not alter
%significantly
%the size of the large leading correction, particularly 
%when Yukawa contributions are present, but we are aware that they
%should be at least qualitatively estimated. This requires 
%the preparation of a complete numerical program at
%one loop where they can be computed, a long work that we 
%made for a process at lepton colliders~\cite{SESAMO} and
%that we are ready to repeat for these different  
%$q\bar q$ and $gg$ initial states. 
With this premise, it seems
to us that a relevant feature that emerges is that the 
relative size of the genuine SUSY correction could be large,
definitely larger that in the SM case, mostly owing to 
the role of the Yukawa correction. 
%In this spirit, we are now
%initiating a more general analysis of production of 
%different final pairs where a SUSY virtual effect could be relevant. A
%priori, we expect such an effect for the production of 
%final squarks, but also for that of charged Higgs and chargino
%pairs, and this general analysis in already in progress.\par

Another comment is related to the possibility that 
the SUSY scenario is not as "reasonably" light as we assumed. In the
case for which the relevant masses are not huge, we feel, 
though, that the SUSY virtual effect, to be computed in a less
simple way i.e. without logarithmic expansions, should 
still be numerically similar to the one that we computed, 
i.e. should not depend dramatically on the values of the SUSY masses of 
the process and should still be computable in a one loop
approximation, and we will devote a future investigation 
to this special (negative) situation.\par 
For what concerns a comparison with other similar work, 
%we are not able at the moment to perform it since, 
%to our knowledge, this paper
%is the first one in the considered analysis of SUSY 
%virtual effects in pair production at proton colliders. 
we remark that our conclusions concerning 
a large virtual effect in the
MSSM are in line (but with SUSY enhanced contributions) 
with those of an analysis of WW 
production in the Standard Model 
framework~\cite{DennerAccomandoPozzorini}. It should
also be recalled that, again within the Standard Model 
framework, the production of b pairs appears to be promising for
the detection of virtual effects in precision 
measurements~\cite{Moretti}. In this sense, we believe 
to have shown that  this
conclusion could still be valid (and even more spectacular)
for a more general 
class of final pairs, in a (hopefully valid) Supersymmetric picture.

\newpage

\begin{figure}
\centering
\vspace{1cm}
\epsfig{file=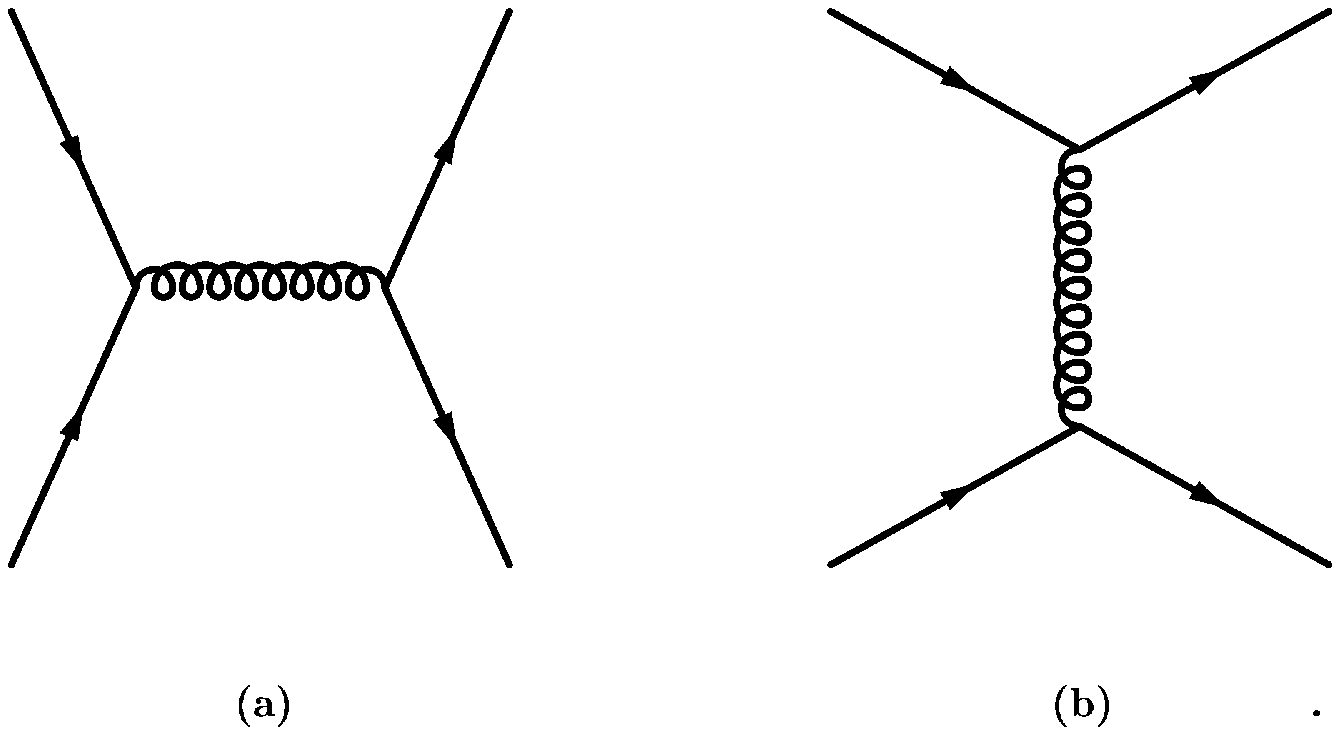,width=12cm}
\vspace{0.5cm}
\caption{Born diagrams for $q\bar q \to q'\bar q'$
annihilation (a) and $q\bar q \to q\bar q$ scattering (b).}
\label{diagram1}
\end{figure}

\newpage

\begin{figure}
\centering
\epsfig{file=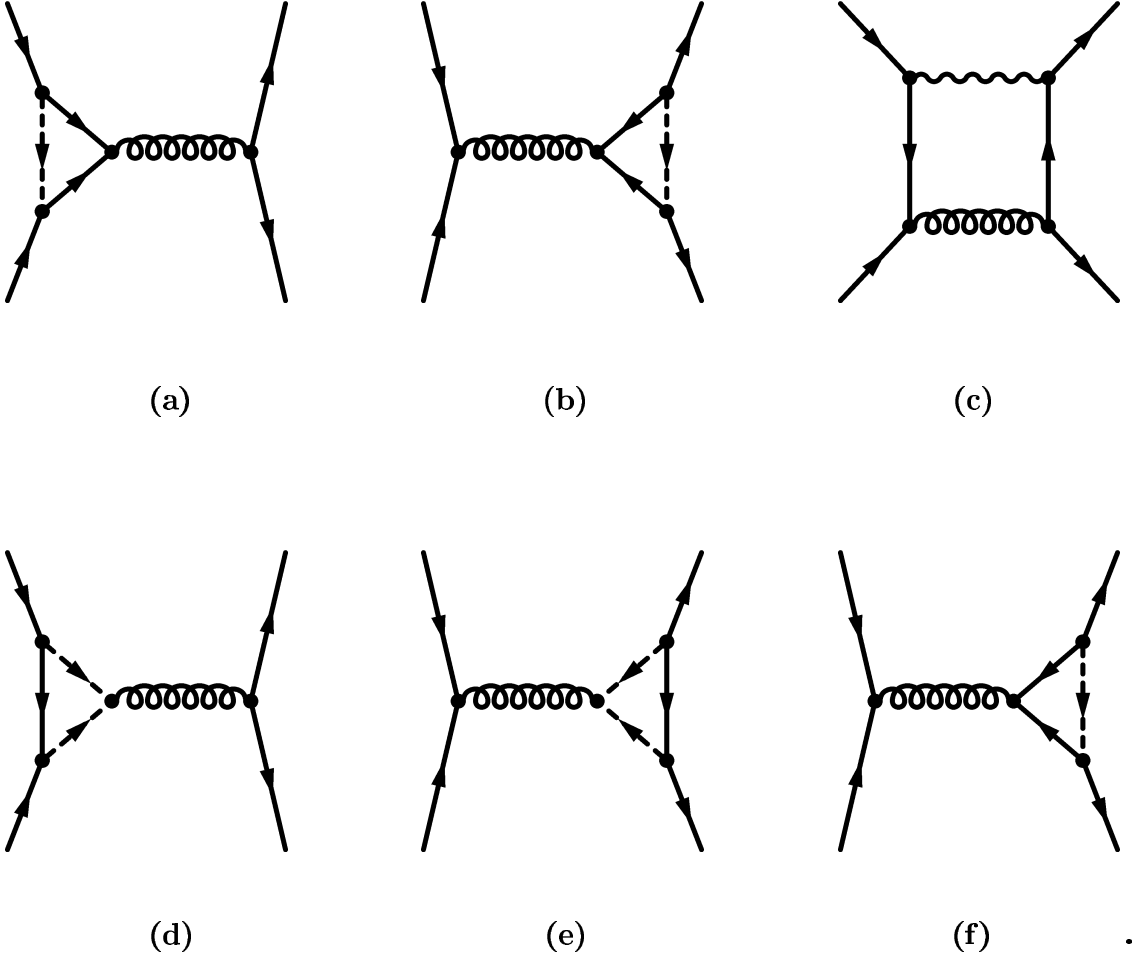,width=15cm}
\vspace{0.5cm}
\caption{Diagrams for electroweak corrections
to the annihilation amplitude $q\bar q \to q'\bar q'$;
Standard Model (a,b,c), where solid lines represent quarks, 
in (a,b) dashed lines represent gauge
or Higgs bosons and in (c) the wavy line is a photon or a $Z$; 
SUSY diagrams (d,e,f), where in (d,e) the internal solid line is a
gaugino and the dashed line is a squark, and in (f) 
solid lines represent quarks and the dashed line is
a SUSY Higgs boson.}
\label{diagram2}
\end{figure}

\newpage

\begin{figure}
\centering
\epsfig{file=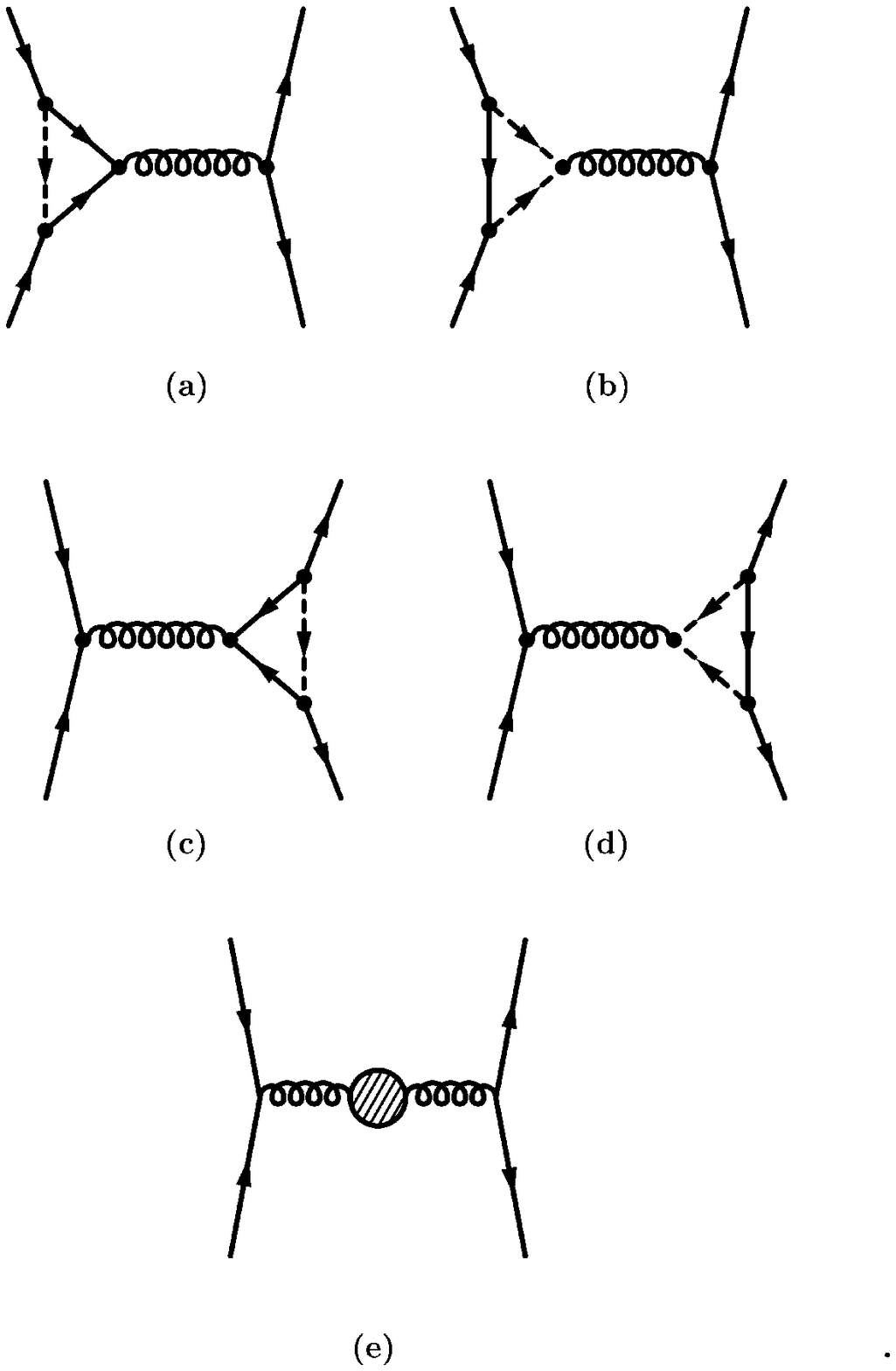,width=12cm}
\vspace{0.5cm}
\caption{Diagrams for SUSY QCD corrections to the
$q\bar q \to q'\bar q'$ annihilation amplitude (a,b,c,d), 
in which the
internal solid lines are gluinos and the dashed lines are squarks,
and diagram for
gluon self-energy diagrams (e).}
\label{diagram3}
\end{figure}

\newpage

\begin{figure}
\centering
\epsfig{file=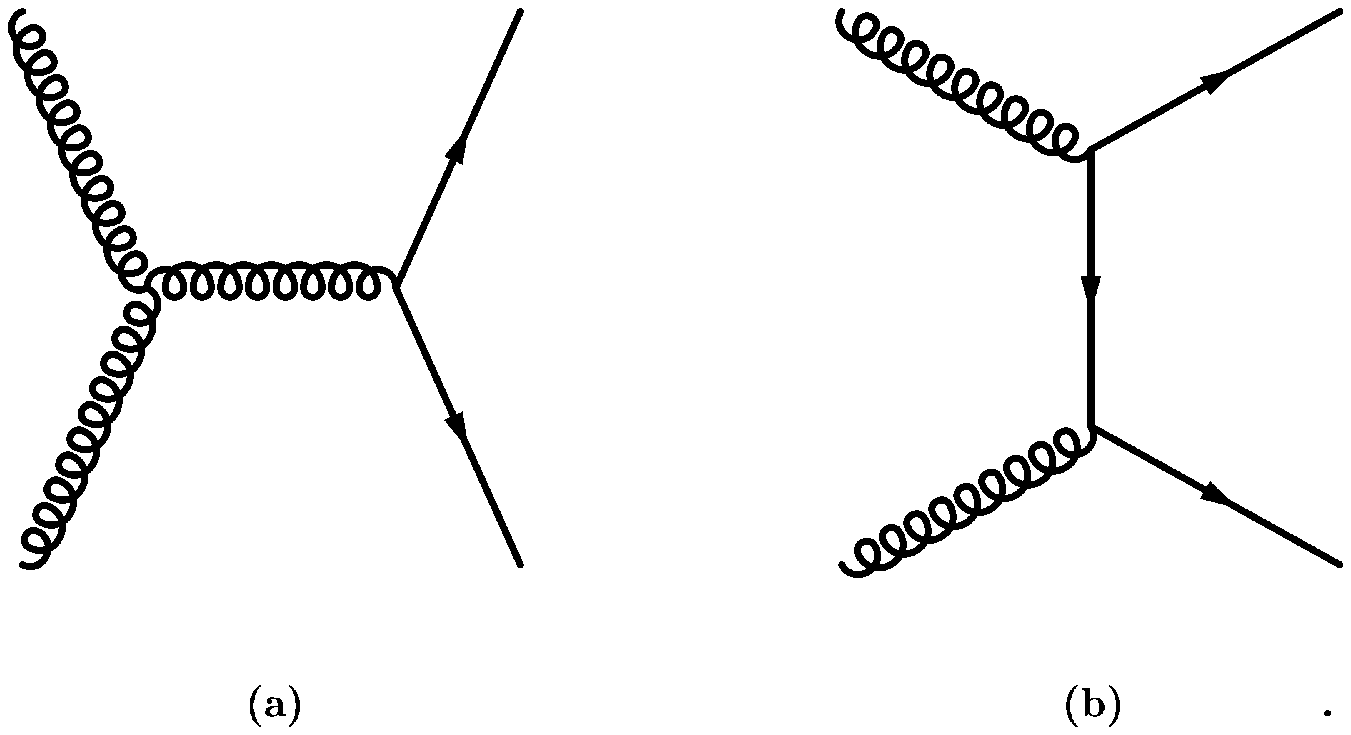,width=12cm}
\vspace{0.5cm}
\caption{Born diagrams for $gg \to q'\bar q'$,
s-channel gluon exchange(a) and t-,u- channel quark exchanges
(b).}
\label{diagram4}
\end{figure}

\newpage

\begin{figure}
\centering
\epsfig{file=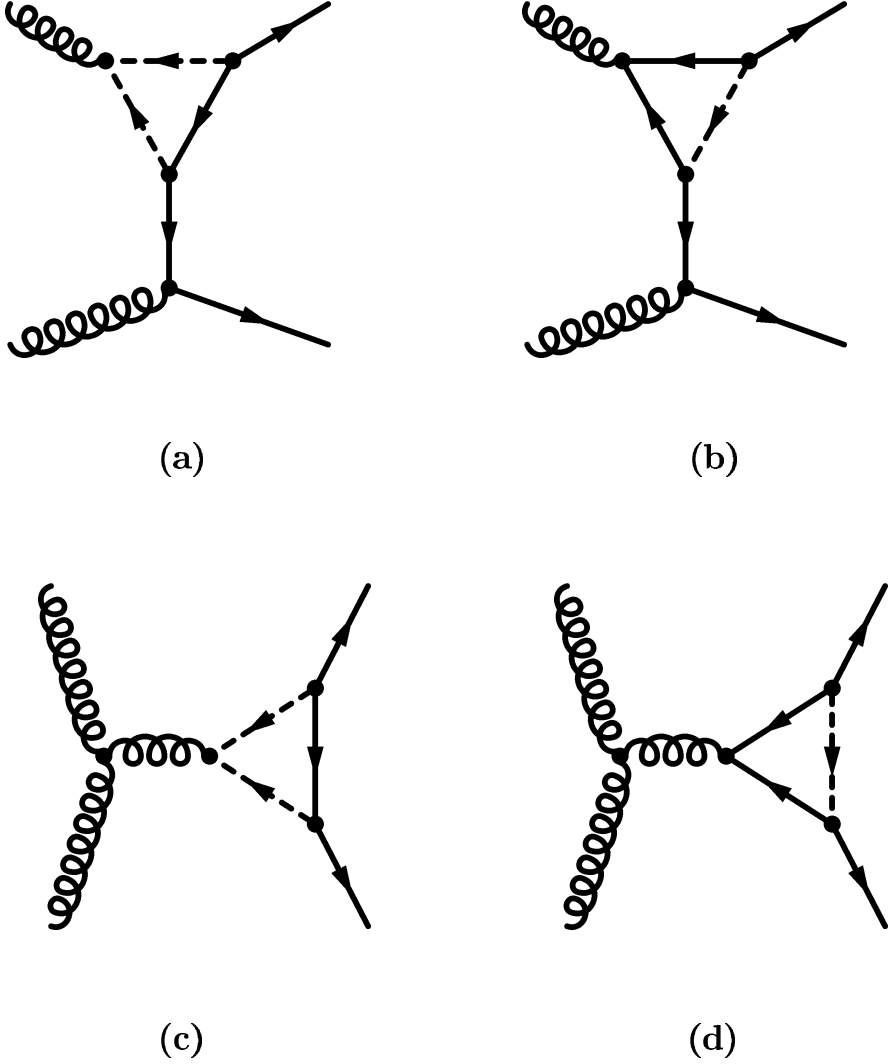,width=12cm}
\vspace{0.5cm}
\caption{Diagrams for SUSY QCD corrections to $gg \to q'\bar q'$
in the t-,u- channels (a,b) (one should add similar diagrams
to (a,b) with down triangles) and in s-channel (c,d); 
in (a,c) the triangles
contains squark (dashed) and gluino (solid) lines; in (b,d) 
they contains
quarks (solid) and SUSY Higgs bosons (dashed) lines.}
\label{diagram5}
\end{figure}

\newpage

\begin{figure}
\centering
\epsfig{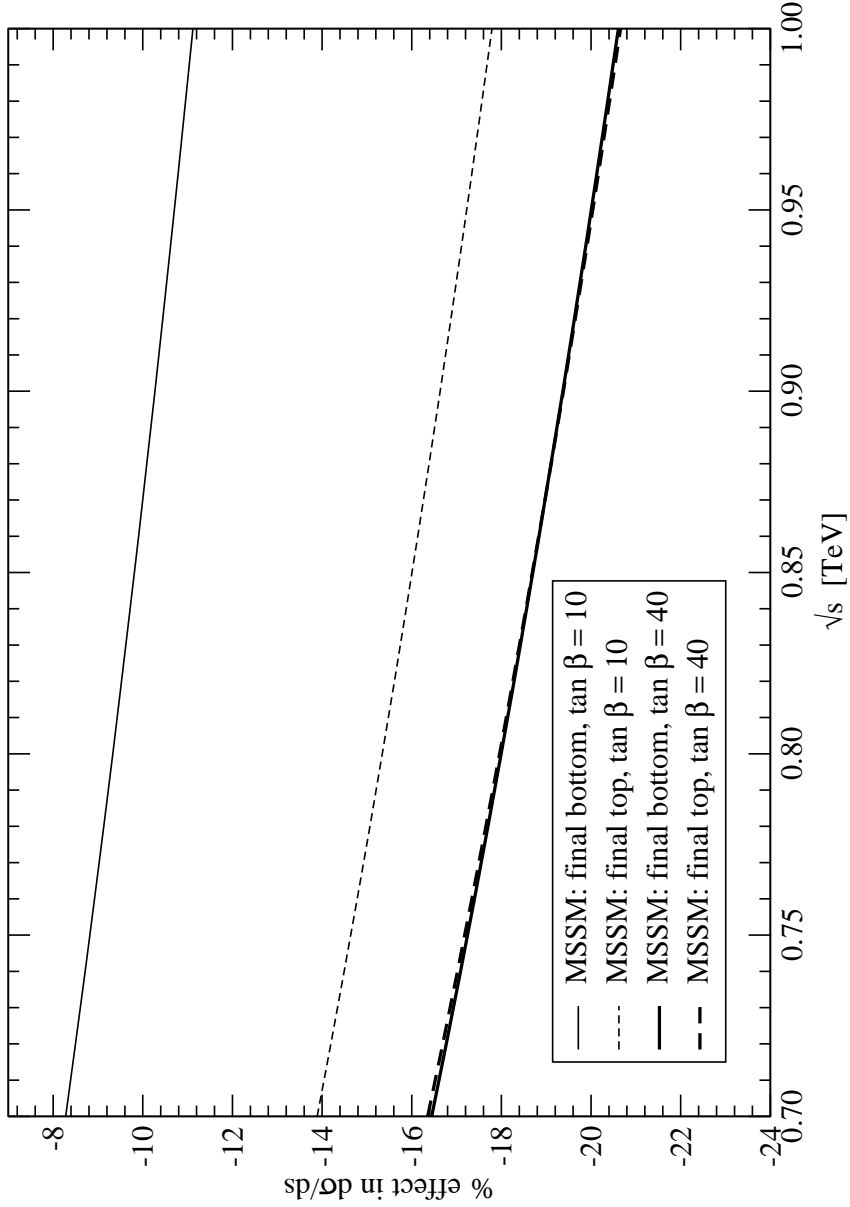}
\vspace{1.5cm}
\caption{Effect of the combined electroweak and SUSY 
QCD corrections in the cross section for final bottom or top pairs
at LHC.
The various parameters are $\sqrt{S}$ = 14 TeV, $M_{SUSY}$ = 350 GeV and $p_{T, min}$ = 10 GeV. We show the results 
obtained with two values of the MSSM parameters $\tan\beta$.}
\label{run1}
\end{figure}

\newpage

\begin{figure}
\centering
\epsfig{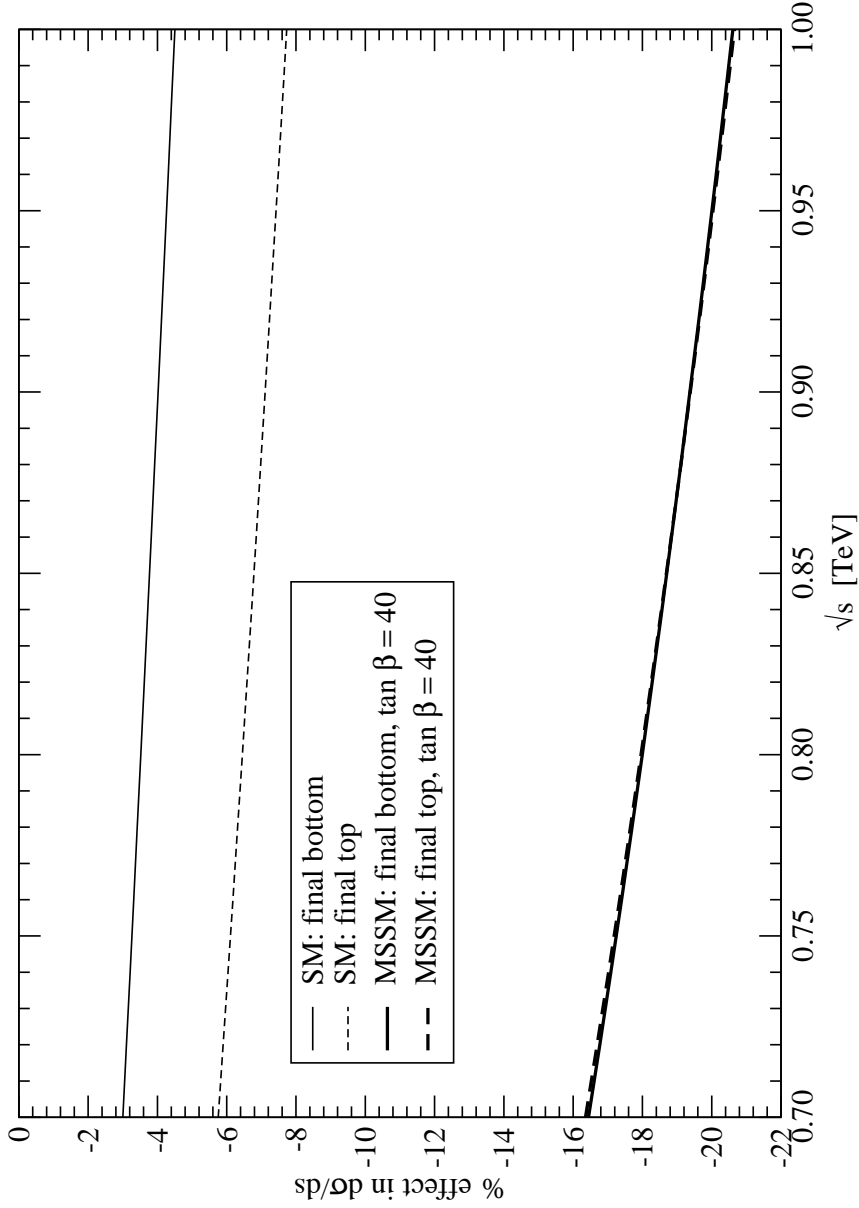}
\vspace{1.5cm}
\caption{Comparison of the full effect in the Standard Model and in the MSSM. We show the correction to the cross section 
for final bottom or top pairs
at LHC in the Standard Model and in the MSSM with $\tan\beta = 40$. 
The other parameters are as in previous Figure: $\sqrt{S}$ = 14 TeV, $M_{SUSY}$ = 350 GeV and $p_{T, min}$ = 10 GeV.}
\label{run2}
\end{figure}

\newpage

\begin{figure}
\centering
\epsfig{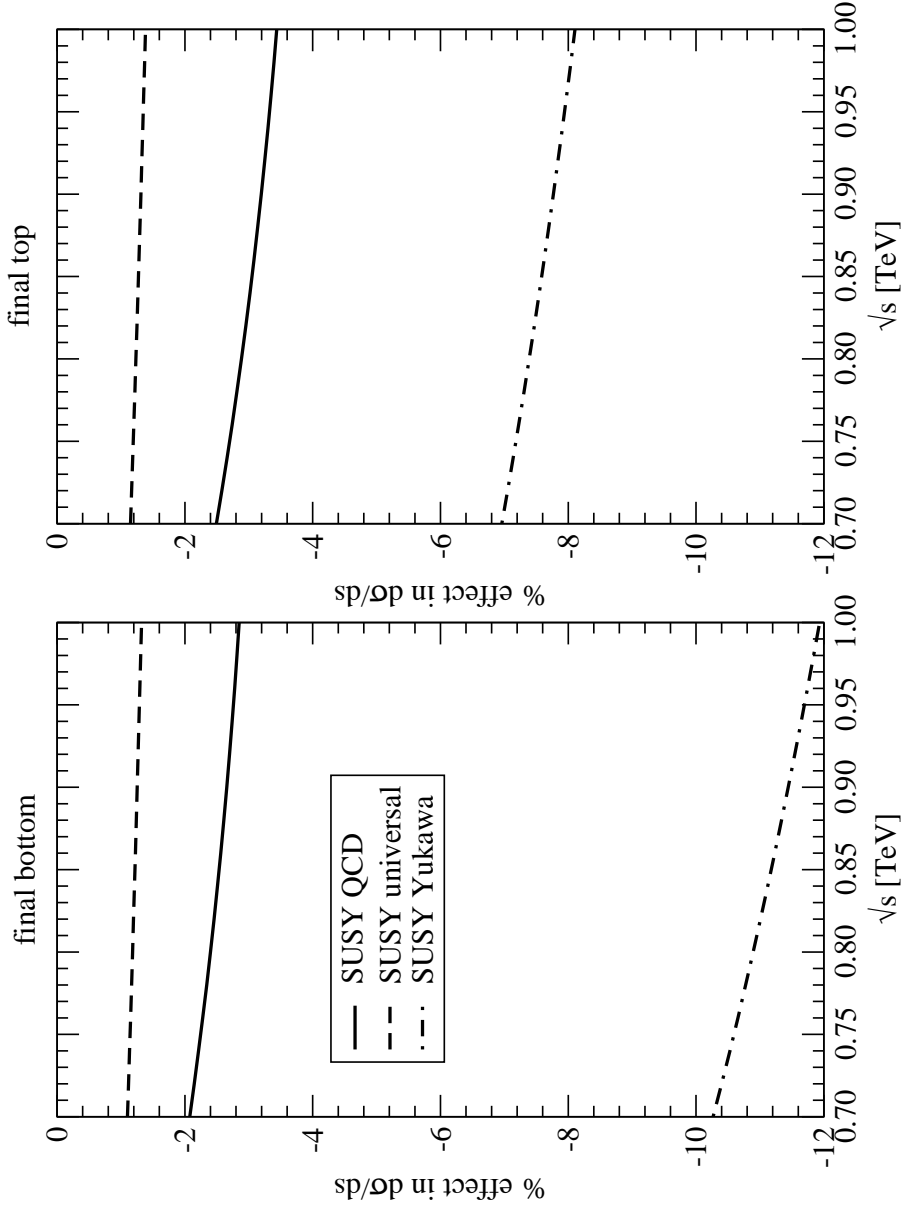}
\vspace{1.5cm}
\caption{Separate SUSY effects. We show the three extra SUSY effects 
that are present in the MSSM with respect to the
Standard Model. They are the SUSY component of the QCD correction, the universal SUSY electroweak terms 
and the $\tan\beta$ dependent SUSY Yukawa terms computed at $\tan\beta = 40$. 
We recall that these genuine SUSY contributions grow like $\log s$. 
The other parameters are $\sqrt{S}$ = 14 TeV, $M_{SUSY}$ = 350 GeV and $p_{T, min}$ = 10 GeV.}
\label{run3}
\end{figure}

\end{document}